# Negative capacitance overcomes Schottky-gate limits in GaN high-electron-mobility transistors

Asir Intisar Khan[1,2†], Jeong-Kyu Kim[3†], Urmita Sikder[1†], Koushik Das[1,2,5], Thomas Rodriguez[3], Rohith Soman[3], Srabanti Chowdhury[3,4], Sayeef Salahuddin[1,2*]

[1]Department of Electrical Engineering and Computer Sciences, University of California, Berkeley, Berkeley, CA, USA

[2]Materials Sciences Division, Lawrence Berkeley National Laboratory, Berkeley, CA, USA

[3]Department of Electrical Engineering, Stanford University, Stanford, CA, USA

[4]Department of Materials Science and Engineering, Stanford University, Stanford, CA, USA

[5]Department of Chemistry, University of California, Berkeley, Berkeley, CA, USA

[†]These authors contributed equally to this work.

[*]E-mail: sayeef@berkeley.edu

**For high-electron-mobility transistors based on two-dimensional electron gas (2DEG) within a quantum well, such as those based on AlGaN/GaN heterostructure, a Schottky-gate is used to maximize the amount of charge that can be induced and thereby the current that can be achieved. However, the Schottky-gate also leads to very high leakage current through the gate electrode. Adding a conventional dielectric layer between the nitride layers and gate metal can reduce leakage; but this comes at the price of a reduced drain current. Here, we used a ferroic HfO₂-ZrO₂ bilayer as the gate dielectric and achieved a simultaneous increase in the ON current and decrease in the leakage current, a combination otherwise not attainable with conventional dielectrics. This approach surpasses the conventional limits of Schottky GaN transistors and provides a new pathway to improve performance in transistors based on 2DEG.**

High-electron-mobility transistors (HEMTs) based on wide bandgap materials are promising for high-frequency and high-power applications (*1–3*). The two-dimensional electron gas (2DEG) at an AlGaN-GaN interface has excellent carrier mobility (*4–6*), which together with the wide bandgap of the nitrides, has led to the adoption of gallium nitride (GaN)-HEMTs for radio-frequency (RF) applications (*7–10*).

Precise heterostructure design and interface engineering are necessary to maintain the charge density of a 2DEG and its mobility. For example, in an AlGaN-GaN heterostructure comprising a GaN channel (2DEG), a barrier (AlGaN) and a capping layer (GaN) are needed; reducing the thickness of the barrier and capping layers can have deleterious effects on the 2D charge density (*11, 12*). For an archetypal HEMT device, a composite dielectric comprising several nanometers of barrier and capping layers is therefore present on top of the transistor channel. Capacitance associated with this composite dielectric is the maximum gate capacitance that can be achieved for a given heterostructure.

Because charge is proportional to capacitance, to ensure the maximum charge and thereby the maximum current, Schottky gated HEMTs have been used where a metal is directly placed on top of the capping layer (*10*). In this configuration, the metal electrode sees the maximum capacitance ($C_{g,sch}$, **Fig. 1A**) achievable for the heterostructure, which ensures achieving the highest current possible. However, leakage current becomes an operational challenge (*13–16*). Although GaN (~3.4 eV) and $Al_xGa_{1-x}N$ (typically ~4.1–4.3 eV in HEMT stack) are considered wide-bandgap semiconductors, their bandgaps and conduction band offsets with respect to the gate metal are lower than those of standard gate dielectrics such as $HfO_2$ (~6 eV bandgap), which leads to substantial leakage through carrier tunneling (*14–16*). Previous studies have attributed such leakage to mechanisms such as Frenkel–Poole emission (*15, 16*) or trap-assisted tunneling (*17*), depending on trap density, barrier thickness, and electric field. A thin layer of a conventional dielectric can exponentially reduce the leakage (*18–21*); but it adds a series capacitance to the $C_{g,sch}$, which reduces the capacitance (*20, 22–25*) seen by the metal electrode (**Fig. 1B**). This approach reduces the amount of charge that can be modulated for the same voltage and therefore the ON current, and leads to a fundamental trade-off between ON and OFF current for the GaN HEMTs.

Inspired by recent demonstrations for silicon transistors showing that a mixed-phase, ferroic $HfO_2$-$ZrO_2$ layer can unconventionally increase the overall capacitance of the series capacitor network of a gate stack (*26–29*) (rather than decrease it) through the negative capacitance (NC) effect (*30–35*), we have used a similar gate stack on a Nitrogen (N) Polar GaN HEMT (**Fig. 1, C and D**). Our results show that, similar to silicon transistors, the overall gate capacitance of the GaN HEMT increased, which made it larger than $C_{g,sch}$. This unconventional increase in capacitance was accompanied by an increase in the ON current, showing that this integration of the ferroic layer did not have any deleterious impact on the transport properties of the electrons in the 2DEG. At the same time, the additional physical thickness of this ferroic layer together with the large bandgap ensured that the leakage current was reduced by more than an order of magnitude compared to the Schottky-gated HEMT (**Fig. 1E**).

### Ferroic-NC dielectric integration and GaN HEMT fabrication

We used a GaN HEMT based on a state-of-the-art AlGaN/GaN heterostructure (schematic in **fig. S1A**). Details of the heterostructure and corresponding device fabrication process are described in Supplementary **Materials and Methods** and **fig. S1 and S2**. The GaN capping layer, AlGaN barrier, and GaN channel layer within the heterostructure are integral to increasing the 2DEG charge density in the channel (induced from back barriers) (*26–29*), which is essential for the functionality of AlGaN/GaN HEMTs (**fig. S3**). We fabricated two different transistors using this heterostructure. For both, the cap layer was recessed down from its initial thickness. For one, a gate metal (tungsten) was deposited directly on the cap layer (schematic shown in **Fig. 1A and 2A**), which we call the Schottky HEMT. For the other, a bilayer of HfO₂-ZrO₂ was first deposited on the recessed cap layer, starting with 13 atomic layer deposition (ALD) growth cycles of ZrO₂ and ending with 5 cycles of ALD grown

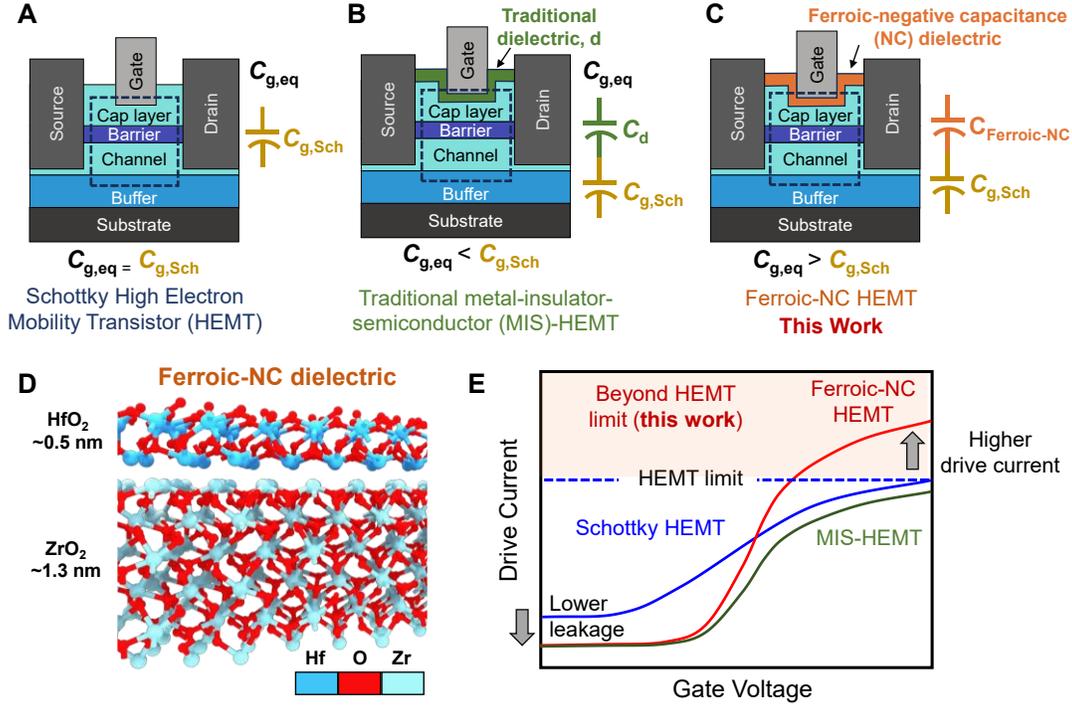

**Fig. 1. Conceptual framework for performance enhancement in GaN HEMTs using ferroic-negative capacitance gate dielectric.** **(A)** Schematic cross-section of a Schottky HEMT where the effective equivalent gate capacitance ($C_{\text{g,eq}}$) corresponds to the total capacitance ($C_{\text{g,Sch}}$) of the underlying gate stack, including contributions from the capping, barrier, and channel layers. The term Schottky refers to the absence of any gate dielectric between the gate metal and recessed GaN capping layer. **(B)** Introduction of a conventional insulating gate dielectric in the gate stack transforms the device into a metal-insulator-semiconductor (MIS) HEMT, which reduced gate leakage but lowered the total gate capacitance below $C_{\text{g,Sch}}$ because of the series capacitance effect. Here the added traditional dielectric contributed a positive capacitance ($C_{\text{d}}$) in series with the intrinsic Schottky gate capacitance $C_{\text{g,Sch}}$, resulting in $C_{\text{g,eq}} = (1/C_{\text{d}} + 1/C_{\text{g,Sch}})^{-1} < C_{\text{g,Sch}}$. The MIS-HEMT structure in **B** is discussed for conceptual framework and was not part of the experimental investigation in this study. **(C)** Incorporation of a ferroic oxide dielectric (*29, 36–38*) into the Schottky configuration could enhance gate capacitance beyond the Schottky limit ($C_{\text{g,Sch}}$) while maintaining low leakage characteristics. The ferroic gate dielectric introduced a negative capacitance (*26, 32–35, 39, 40*) effect ($C_{\text{Ferroic-NC}}$) to $C_{\text{g,Sch}}$ resulting in $C_{\text{g,eq}} > C_{\text{g,Sch}}$ (further discussion in **Fig. 4**). **(D)** Schematic of the ferroic-NC dielectric used in this work, comprising an ultrathin bilayer of ZrO₂ (~1.3 nm)/HfO₂ (~0.5 nm) referred to as ~1.8 nm HZO. **(E)** Schematic representation of the drive current-gate voltage transfer characteristics comparing ferroic-NC HEMT, Schottky HEMT, and MIS-HEMT. The NC HEMT leveraged enhanced capacitance and gate control to achieve higher ON-state drive current, surpassing the Schottky HEMT limit, while simultaneously reducing OFF-state leakage.

HfO$_2$, capped in-situ in ALD with metal TiN. Tungsten was deposited on this bilayer as the gate metal. We call this device the NC HEMT (**Fig. 1C and 2C**).

To ensure that both transistors had the same thickness of the cap layer after recessing, both samples were etched together at the same time. Additionally, atomic force microscopy (AFM) was performed immediately after recess etching had been performed. The imaging confirmed that the recessed regions for both structures had the same trench depth (**fig. S4**). This finding was further corroborated by post-fabrication transmission electron microscopy (TEM) studies that we discuss below.

**Structural Characterization**

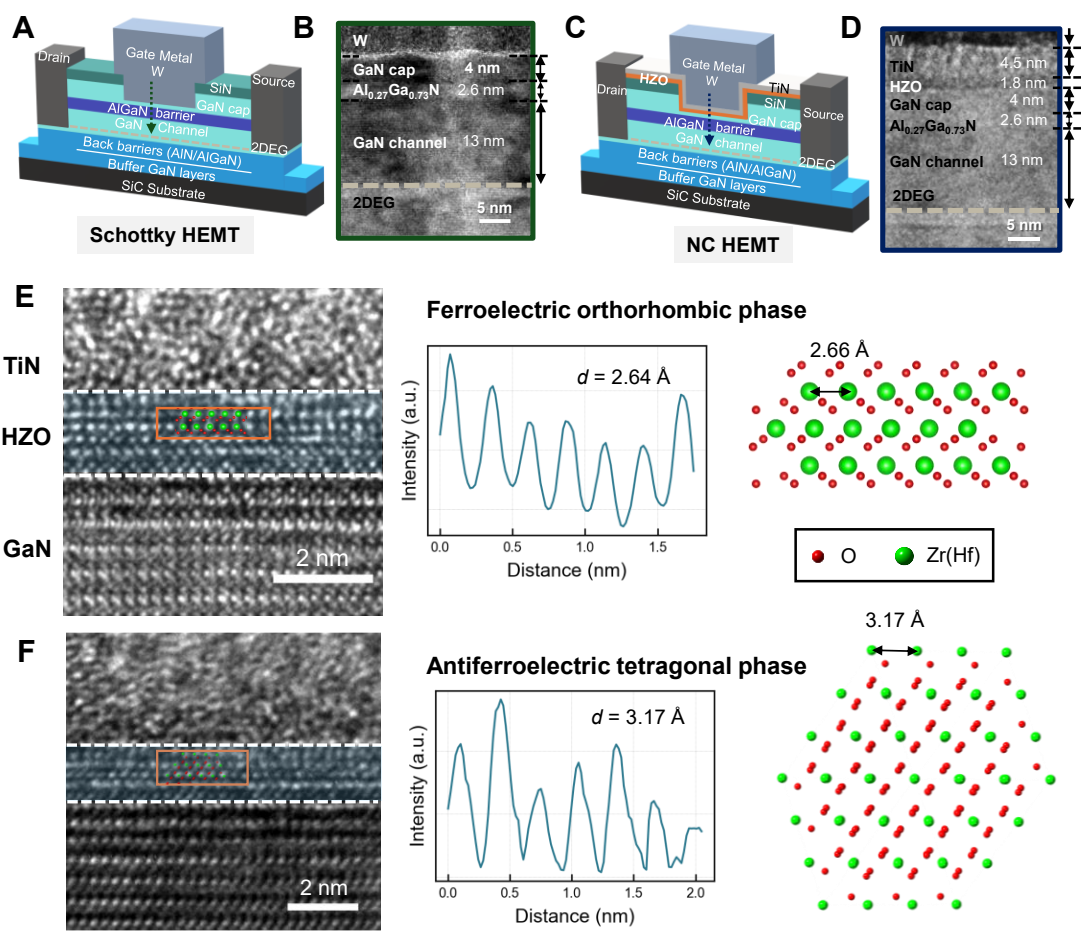

**Fig. 2. Schottky and NC GaN HEMT structures and materials characterization. (A)** Schematic of the fabricated Schottky GaN HEMT. The term Schottky refers to the absence of any gate dielectric between the gate metal and recessed GaN capping layer. **(B)** TEM cross-sectional image of the Schottky HEMT stack beneath the tungsten (W) gate metal (cutline shown in dotted arrow in **A**). It reveals a ~4 nm thick recessed GaN cap layer. Energy-dispersive spectroscopy (EDS) mapping of the Schottky HEMT layers is presented in **fig. S5A**. **(C)** Schematic representation of the fabricated NC GaN HEMT with ferroic-NC HZO dielectric (capped in-situ with metal TiN) between the gate metal (W) and recessed GaN capping layer. **(D)** TEM cross-sectional micrograph of the NC HEMT beneath the gate metal (cutline shown in dotted arrow in **C**), showing the presence of a ~1.8 nm HZO gate dielectric on a ~4 nm remaining GaN cap layer which is identical in thickness to that of the Schottky GaN HEMT. (EDS mapping of NC HEMT stack is shown

in **fig. S5B**). Both devices also show identical nitride epitaxial layers, including a ~13 nm GaN channel layer and a ~2.6 nm $Al_{0.27}Ga_{0.73}N$ barrier layer. Specifications of the full stacks, including back barriers and buffer layers, are described in **fig. S1 and fig. S2**. High-resolution TEM (HRTEM) cross-sectional images of the atomic-scale 1.8 nm thick HZO dielectric and extracted $d$-lattice spacings (determined from the orange box regions) showing the simultaneous presence of **(E)** fluorite-structure ferroelectric orthorhombic $(Pca2_1)$ and **(F)** antiferroelectric tetragonal $(P4_2/nmc)$ phases in the same HZO layer, promoting the stabilization of the negative capacitance effect. Corresponding zoomed-out TEM and EDS analysis of the stack in **E**, **F** are presented in **fig. S6A**. The HZO layer in **E** and **F** was directly deposited using the identical condition as for the devices. The phases were assigned based on the atomic configuration of Zr(Hf) and their interatomic spacings. Atomic configuration of ferroelectric orthorhombic $(Pca2_1)$ phase along [100] zone axis, and antiferroelectric tetragonal $(P4_2/nmc)$ phase along [111] zone axis are presented in **E** and **F**, respectively. Detailed comparisons of the atomic configurations of different phases with similar d-spacings are shown in **fig. S6B**.

TEM was used to image the cross-section of both the Schottky HEMT (**Fig. 2B**) and the NC HEMT (**Fig. 2D**). A ~1.8 nm thin dielectric layer is clearly visible in **the** NC HEMT (**Fig. 2D**). The thickness is expected from the 18 ALD cycles (13 for $ZrO_2$ and 5 for $HfO_2$) used for synthesis of this bilayer. The TEM images (**Fig. 2, B and D**) were consistent with atomic force microscopy (AFM) measurements (**fig. S4, A and B**), further confirming the identical thicknesses of the different layers: recess etched GaN cap (~4 nm), $Al_{0.27}Ga_{0.73}N$ (~2.6 nm), and GaN channel (~13 nm) in both devices. Recess-etched regions of both devices showed low surface roughness with a root-mean square roughness of ~0.6 to ~0.7 nm (**fig. S4, C and D**). The energy dispersive spectra (EDS) analysis confirmed the elemental distribution and their uniformity in each layer of the stacks for both types of HEMTs (**fig. S5**), confirming that the only difference between the fabricated devices is the presence of a ~1.8 nm HZO dielectric capped in-situ with ~4.5 nm metal TiN in the NC HEMT. Additional HRTEM images indicated the co-existence of the ferroelectric (FE) orthorhombic (**Fig. 2E**) and antiferroelectric (AFE) tetragonal phases (**Fig. 2F**) in our fluorite-structure HZO film deposited on GaN. The presence of mixed FE–AFE behavior within the same HZO film was conducive to stabilizing the negative capacitance effect (*26–29*, *41*, *42*) and will be discussed below.

**Electrical Measurements**

An optical micrograph and SEM image of a representative device are shown in **Fig. 3, A** and **B**, respectively. The direct current (d.c.) transfer characteristics ($I_D$ vs. $V_{GS}$) showed a higher maximum drive current in the NC-HEMT compared to the Schottky devices with the same dimensions (**Fig. 3C** and **fig. S7, A and B**) and across various drain voltages (**fig. S8A**; see Supplementary **Materials and Methods** for electrical measurement method). Additionally, the NC-HEMTs exhibited more than an order of magnitude lower leakage current, which we expected from the addition of the larger bandgap, $HfO_2$-$ZrO_2$ bilayer as compared to Schottky devices that have no such dielectric beneath the gate metal. Notably, the higher drive current was also evident when the currents are compared for the same overdrive ($V_{GS}$-$V_{th}$) of the respective HEMT devices (**fig. S8B**). Furthermore, the NC-HEMT exhibited nearly hysteresis-free transfer characteristics (bidirectional $I_D$–$V_{GS}$ sweeps in **fig. S9**), indicating a high-quality dielectric-GaN cap interface.

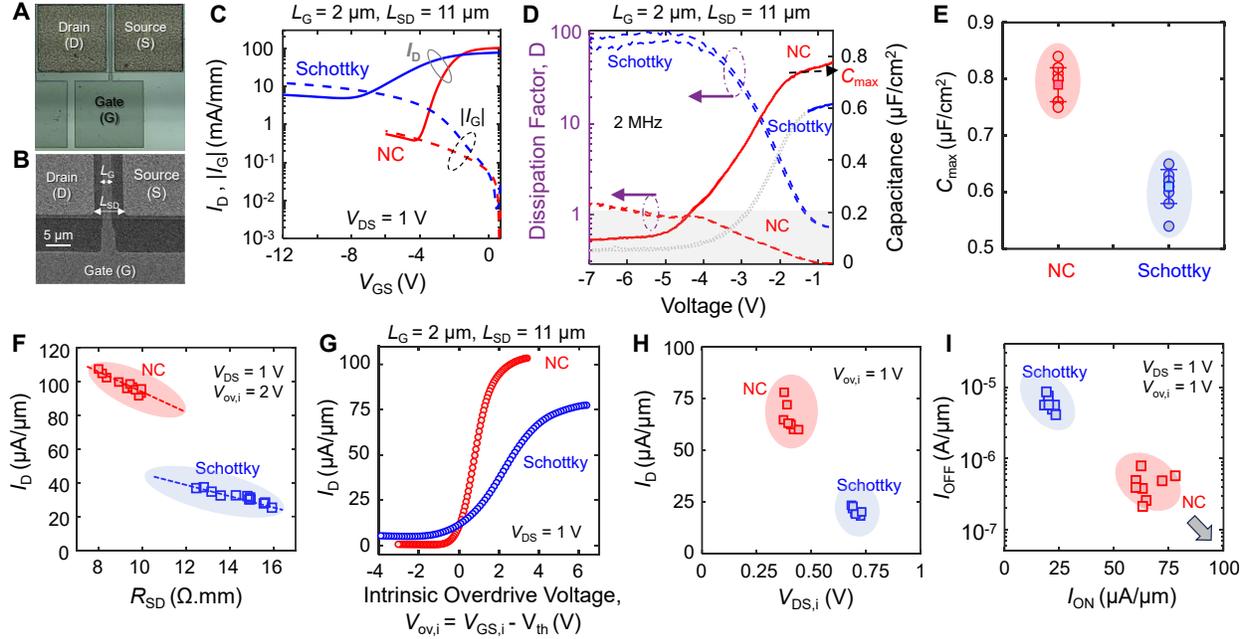

**Fig. 3. Electrical performance of NC- and Schottky GaN HEMTs. (A)** Optical micrograph and **(B)** top-view SEM image of a fabricated HEMT device. $L_{SD}$ is the source-to-drain distance, and $L_G$ is the gate length. All the devices have an $L_G$ of 2 µm, varying $L_{SD}$, and channel width $W$ of 100 µm. **(C)** Measured d.c. transfer characteristics: drain current, $I_D$ vs. gate voltage $V_{GS}$ for a HEMT with ferroic-NC HZO dielectric (solid red line) and Schottky-gated HEMT (solid blue line), both with $L_{SD}$ = 11 µm. Gate leakage currents ($|I_G|$) are shown in dashed lines. $I_D$ was normalized by $W$ and measured at drain voltage $V_{DS}$ = 1 V. **(D)** Representative capacitance and dissipation factor (D) as functions of gate voltage at 2 MHz for Schottky (blue) and NC (red) HEMTs ($L_G$ = 2 µm, $L_{SD}$ = 11 µm) measured at a low $V_{DS}$ of 10 mV. D is defined as the ratio of the equivalent series resistance to capacitive reactance. The NC device exhibits a higher maximum gate capacitance ($C_{max}$) compared to Schottky HEMT, indicating enhanced gate-channel coupling due to the amplification effect of the NC dielectric. The dotted gray portions of the $C$-$V$ curves (D > 1), reflect large leakage and unstable $C$-$V$ curve region. In contrast, the solid portions of the curves represent regions where D < 1 and the capacitance data are reliable. **(E)** Measured maximum gate capacitance, $C_{max}$ at 2 MHz for NC GaN HEMTs compared to Schottky HEMTs. Square symbols and error bars mark the average and SD, respectively, across seven devices (circles) of each type. **(F)** $I_D$ vs. $R_{SD}$ scatter plot at $V_{DS}$ = 1 V and $V_{ov,i}$ = 2 V for ferroic-NC and Schottky HEMTs. $R_{SD}$ is the sum of source resistance (source contact resistance + source access region resistance) and drain resistance (drain contact resistance + drain access region resistance) [extracted from **fig. S10A**]. $V_{ov,i}$ is the intrinsic overdrive voltage defined as $V_{ov,i}$ = $V_{GS,i}$ − $V_{th}$ where $V_{GS,i}$ = $V_{GS}$ − $I_D R_{SD}/2$. Dashed line trends (guide to the eye) display that NC-HEMT devices achieved higher $I_D$ for the same $R_{SD}$. **(G)** $I_D$ vs. $V_{ov,i}$ (at $V_{DS}$ = 1 V) confirmed higher $I_D$ in NC-HEMT at the same $V_{ov,i}$. **(H)** $I_D$ vs. intrinsic drain voltage $V_{DS,i}$ (at $V_{ov,i}$ = 1 V) scatter plot revealed higher drive current at a lower $V_{DS,i}$ for NC-HEMTs compared to those of Schottky devices. $V_{DS,i}$ = $V_{DS}$ − $I_D R_{SD}$. **(I)** On-state current $I_{ON}$ (defined as the drain current at $V_{DS}$ = 1 V, $V_{ov,i}$ = 1 V) vs. off-state current $I_{OFF}$ demonstrated simultaneously higher drive current and lower leakage current for HEMTs with ferroic-NC HZO dielectric, overcoming the fundamental drive current limit of control Schottky HEMT. The block arrow points to the desirable "best corner" with low $I_{OFF}$ and high $I_{ON}$.

Capacitance vs. voltage characteristics are shown in **Fig. 3D**. For the Schottky-HEMT, we can identify a voltage window of ~1 V (from approximately -1.5 V to -0.5 V) where the dissipation factor, D was small enough (D < 1), indicating low enough leakage current, and enabling high fidelity capacitance measurement (**Fig. 3D**). This low-leakage regime corresponds to low gate biases where the electric field across the gate stack is not strong enough to trigger significant tunneling or barrier lowering. Within this entire region, the NC-HEMT had a substantially larger capacitance than the Schottky-HEMT. On average, the measured gate capacitance for the NC-HEMT devices was ~30% higher than that in the Schottky-HEMT (**Fig. 3E**).

The measured drive current $I_D$ of a transistor could be influenced or limited by the source-to-drain series resistance ($R_{SD}$) including source-drain contact resistance and access region resistance. To evaluate the intrinsic $I_D$, we extracted $R_{SD}$ of both NC and Schottky HEMTs (**fig. S10A**). The average $R_{SD}$ in our Schottky HEMTs was higher than that in NC HEMTs, primarily because of a higher source-drain contact resistance in the Schottky devices (**fig. S10B**). Nevertheless, the NC devices exhibited higher intrinsic $I_D$ at the same projected $R_{SD}$; $I_D$ measured at an intrinsic overdrive voltage ($V_{ov,i}$) of 2 V and $V_{DS} = 1$ V (**Fig. 3F**). The plots of $I_D$ vs. $V_{ov,i}$ (**Fig. 3G**) and $I_D$ vs. intrinsic drain voltage ($V_{DS,i}$) (**Fig. 3H**) further confirmed the consistently higher intrinsic drive current for NC HEMTs compared to their Schottky counterparts. Finally, **Fig. 3I** shows $I_{ON}$-$I_{OFF}$ tradeoff space from the measured data. An almost factor of 3 increase in the ON current and more than an order of magnitude reduction in the OFF current were observed.

**Mechanism of the capacitance changes and ON current increase**

As the structural characterizations (TEM and AFM) show, the only difference between the two HEMT structures is that the NC-HEMT has an additional dielectric and a metal capping layer on top of the recessed GaN cap layer. The reduction in the leakage current for the NC-HEMT was expected because of the additional thickness added to the gate stack. According to conventional electrostatics, this layer should have added a series capacitance to the Schottky HEMT gate capacitance, lowering the overall capacitance for the NC-HEMT and thereby lowering the amount of charge and current that can be modulated. We considered what effects enabled the observed increase in the ON current at the same overdrive voltage.

From electrical measurements, we observed that the NC-HEMT did not follow the conventional electrostatics in that its gate capacitance was larger rather than being smaller compared to that of Schottky-HEMT (**Fig. 3E**). This unconventional increase in the gate capacitance can readily be explained by recognizing that the HfO$_2$-ZrO$_2$ bilayer provides a negative capacitance to the overall stack. A negative capacitance in series with a positive capacitance will increase the overall capacitance.

To self-consistently characterize the specific HfO₂-ZrO₂ bilayer used in this work, we also deposited the same stack on a Si wafer with chemically grown SiO₂ interfacial layer (see Supplementary **Materials and Methods**). In **Fig. 4**, we show that the bilayer provided a similar increase in capacitance as seen before (*26–29*), establishing the action of the HfO₂-ZrO₂ bilayer as negative capacitance. The capacitance boost in ~1.8 nm HZO can be attributed to the negative capacitance effect (**Fig. 4A**) of the ferroelectric HZO layer, which amplifies gate capacitance. Conversely, the control 1.8 nm HfO₂ layer in the SiO₂+HfO₂ stack adds positive capacitance to the system, resulting in a decrease in total gate capacitance below that of the 8 Å EOT SiO₂ alone (**Fig. 4B**). This behavior highlights the unique role of the ferroic-HZO bilayer in improving gate control and overall device performance.

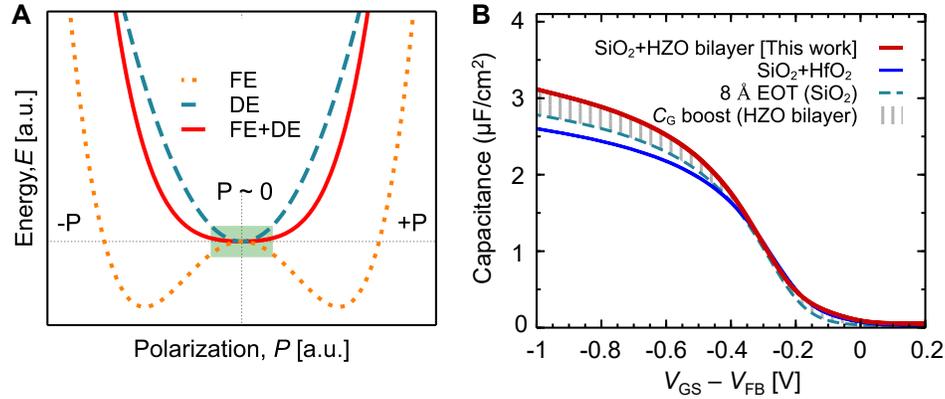

**Fig. 4. Negative capacitance (NC), and capacitance boost in ferroic HfO₂-ZrO₂ bilayer (HZO). (A)** NC effect. Conceptual energy ($E$) vs. polarization ($P$) curves for a ferroelectric (FE, dotted orange), dielectric (DE, dashed green), and combined FE–DE stack (solid red). FE materials exhibit a double-well energy profile due to spontaneous polarization, while conventional dielectrics show a single parabolic well. In the central region of the FE curve, the slope becomes negative, implying that increasing polarization requires decreasing electric field, which corresponds to a negative capacitance region. Although unstable in standalone FE materials, this regime can be stabilized by placing the FE in series with a linear DE, which flattens the overall energy landscape P ~ 0 (shaded green). This stabilization enables the same amount of charge to be achieved with less applied voltage, effectively boosting the total gate capacitance (*31–35*). **(B)** Capacitance-voltage (*C-V*) characteristics for metal–oxide–semiconductor stacks on silicon, each with an identical 8 Å effective oxide thickness (EOT) chemically grown SiO₂ layer: SiO₂ alone (dashed green) (*29*), SiO₂ + 1.8 nm HfO₂ control (blue), and SiO₂ + 1.8 nm HZO bilayer [ZrO₂ (1.3 nm)/HfO₂ (0.5 nm), in red]. Our ~1.8 nm ferroic-HZO bilayer exhibits a capacitance boost relative to both controls, consistent with NC behavior. HZO thickness and deposition process (see **Materials and Methods**) were identical as for the HZO integrated onto GaN HEMT in our study. The *x*-axis, $V_{GS}$–$V_{FB}$ represents the gate voltage relative to the flat-band voltage for consistency across the stacks.

Additionally, various combinations of HfO₂ and ZrO₂ layers on Si have recently been shown to have a mixed-phase ferroic order that leads to negative capacitance behavior (*26–29*). From cross-sectional HRTEM images, we confirmed the presence of mixed phases (FE polar orthorhombic and AFE non-polar tetragonal) within the HfO₂-ZrO₂ bilayer (**Fig. 2, E and F**) grown directly on GaN.

Structural inhomogeneity in mixed phase ferroic films can induce depolarization fields that suppress long-range polarization (*43*). In systems with polar–nonpolar phase coexistence, these fields can stabilize ferroelectric regions in locally higher energy configurations, enabling access to and stabilization of negative capacitance (*31–33, 44*). This is consistent with the behavior observed in our ultrathin bilayer HZO films. Such mixed phase HZO structures have also been shown to maintain low leakage even on ultrathin $SiO_2$ interfacial layers (~8.0 to 8.5 Å EOT) (*26*), underscoring the high quality and effectiveness of such films as the gate dielectric. In our case, the ~1.8 nm HZO layer was deposited on a 4 nm GaN cap layer. This configuration further increased the effective barrier thickness between the gate and the 2DEG, reducing leakage through the combined effects of the large bandgap of HZO and its thickness, which increases the physical separation between the gate metal and the channel.

We also note that unlike in Si-channel transistors, where the gate dielectric interfaces directly with the Si channel through an ultrathin $SiO_2$ interfacial layer, the gate dielectric in our N-polar GaN HEMT is separated from the 2DEG via a 4 nm GaN cap and a ~2.6 nm AlGaN barrier. As a result, the HZO gate dielectric is physically separated from the conduction channel, and interface traps at the dielectric boundary are not expected to significantly affect channel behavior.

## Discussion

High leakage current in Schottky-HEMT has remained a longstanding problem in GaN HEMTs. Classical electrostatics does not provide a plausible path towards reducing leakage without affecting the amount of charge and current that can be modulated with a given overdrive voltage. In this context, mixed phase ferroic-negative capacitance dielectrics could provide completely new pathways for GaN HEMT design. As we demonstrated, such a dielectric could not only reduce leakage current by orders of magnitude, but it could also increase the ON current simultaneously. Notably, the bilayer used in this work was optimized for Si transistors. Focused optimization of NC integration for GaN HEMTs, particularly in scaled or high-frequency designs, could further enhance the performance gains demonstrated here. While our study focuses on N-polar GaN HEMTs, implementing NC gate stacks in Ga-polar devices should be possible through similar recess etching or epitaxial engineering to enhance NC behavior while preserving 2DEG confinement. In addition, recent demonstrations of nitride-based ferroelectrics such as AlScN (*45, 46*) may make it possible to design even stronger negative capacitance by utilizing epitaxial pathways. The key concept also applies generally to improve performance of any transistor with a heterostructure quantum well induced 2DEG such as III-V transistors.

**Acknowledgment:**

A.I.K. is thankful to M. Noshin for useful discussions on GaN heterostructures.

**Funding:** This work was funded primarily by the Berkeley Center for Negative Capacitance Transistors. It was also funded in part by AirForce Office of Scientific Research Grant No: FA9550-25-1-0013. Work at Stanford was funded in part by the System X Alliance.

**Author contributions:** S.S., S.C., and A.I.K. conceived the idea. A.I.K. and S.S. designed and led the research with input from S.C. The device fabrication, electrical measurements and materials characterization were performed by J.-K.K., A.I.K and U.S., with input from T.R., R.S., S.S., and S.C. TEM analysis was performed by K.D. with input from S.S. The manuscript was written by A.I.K., S.S., J.-K.K., and U.S. The research was supervised by S.S. and S.C. All authors contributed to discussions and manuscript preparations.

**Competing interests:** The authors declare no competing interests.

**Data and materials availability:** All data needed to evaluate the conclusions in this paper are present in the paper or the supplementary materials.

**Supplementary Materials**

Materials and Methods

Figs. S1 to S10

References

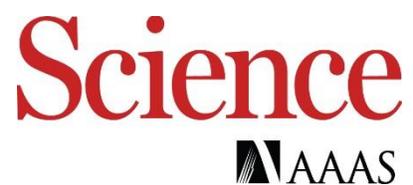

Supplementary Materials for

# Negative capacitance overcomes Schottky-gate limits in GaN high-electron-mobility transistors

Asir Intisar Khan[1,2†], Jeong-Kyu Kim[3†], Urmita Sikder[1†], Koushik Das[1,2,5], Thomas Rodriguez[3], Rohith Soman[3], Srabanti Chowdhury[3,4], Sayeef Salahuddin[1,2*]

[1]Department of Electrical Engineering and Computer Sciences, University of California, Berkeley, Berkeley, CA, USA

[2]Materials Sciences Division, Lawrence Berkeley National Laboratory, Berkeley, CA, USA

[3]Department of Electrical Engineering, Stanford University, Stanford, CA, USA

[4]Department of Materials Science and Engineering, Stanford University, Stanford, CA, USA

[5]Department of Chemistry, University of California, Berkeley, Berkeley, CA, USA

[†]These authors contributed equally to this work.

[*]E-mail: sayeef@berkeley.edu

**The PDF file includes:**

Materials and Methods
Figs. S1 to S10
References

**Materials and Methods**

**GaN HEMT fabrication**. We start with the metal-organic chemical vapor deposition (MOCVD) grown Nitrogen (N)-Polar AlGaN/GaN heterostructure stack (**fig. S1A, fig. S2A**). The growth process of this stack is detailed elsewhere (*47, 48*). A ~400 nm-thick mesa etching was performed using $BCl_3/Cl_2$ based inductively coupled plasma reactive ion etching (ICP-RIE) to effectively isolate the device active region on the substrate. Subsequently, Ti (20 nm) /Al (120 nm) /Ni (30 nm) /Pt (60 nm) source (S)/drain (D) metal contacts were deposited via e-beam evaporation. While a 60 nm of gold (Au) is typically used as a capping layer, platinum (Pt) was used in this work to meet the gold-free conditions required for the subsequent HZO dielectric deposition using atomic layer deposition (ALD). After lift-off process using a resist stripper (Remover PG), a rapid thermal annealing (RTA) at 850 °C was conducted to form alloyed ohmic S/D contacts. Next, gate recess etching was conducted. First, a 7 nm MOCVD $SiN_x$ was etched using $CF_4$-based ICP-RIE. This was followed by $BCl_3/Cl_2$-based ICP-RIE for GaN cap layer etching. Before the HZO deposition, a dilute HF treatment was performed to achieve a cleaner interface in the gate recess region.

A ~1.8 nm HZO was subsequently deposited using thermal ALD, followed by 4 nm of TiN capping layer deposited in-situ via plasma-enhanced ALD. The HZO (1.3 nm $ZrO_2$ / 0.5 nm $HfO_2$) was grown in Cambridge Fiji 200 Plasma ALD system at 300 °C. Tetrakis(dimethylamido)hafnium and Tetrakis(dimethylamido)zirconium, heated to 80 °C, were used as precursors. For the HZO deposition thermal process was utilized, using water vapor as the oxidizing agent. TiN was deposited on the HZO bilayer without breaking the vacuum using a plasma enhanced ALD process. Tetrakis(dimethylamido)titanium was used to deposit 60 cycles of TiN at 300 °C.

Finally, a 150-nm-thick W layer was deposited as the gate metal using direct current (d.c.) sputtering. ICP-RIE was utilized to etch W in all areas except the gate fingers and pad regions. The control Schottky GaN HEMT was fabricated simultaneously with the NC GaN HEMT, except for the gate dielectric ALD steps. The fabrication process is further detailed in **fig. S1 and fig. S2**.

**Si MOS capacitor fabrication.** In order to evaluate the ferroic NC bilayer HZO gate oxide, silicon channel MOS capacitor structures were fabricated. The lightly doped *p*-type silicon samples were prepared by cleaning the wafer with piranha and aqueous HF solution to remove the native oxide. The sample was then soaked in standard clean SC-1 solution, i.e., 5:1:1 $H_2O$:$H_2O_2$:$NH_4OH$ bath at 80 °C for 10 minutes which chemically grows 8-8.5Å $SiO_2$ interfacial layer in a self-limiting process. HZO (bilayer $ZrO_2$ / $HfO_2$) was grown by ALD in a Cambridge Fiji 200 Plasma ALD system at 300 °C. Tetrakis(dimethylamido)hafnium and Tetrakis(dimethylamido)zirconium, heated to 80 °C, were used as precursors. The HZO bilayer was grown in a thermal process, using water vapor as the oxidizing agent. TiN was deposited on the HZO bilayer without breaking the vacuum using a plasma enhanced ALD process. Tetrakis(dimethylamido)titanium was used to deposit 60 cycles of TiN at 300 °C. The deposition process of HZO and TiN was identical to the process used for the GaN HEMTs. 70 nm of W was deposited subsequently by sputtering. MOS capacitor electrodes were defined by photolithography and dry etching in an ICP etcher.

**Negative capacitance (NC) gate stack considerations.** The bilayer HZO stack was deposited using thermal ALD cycles, where the metal precursor was pulsed first, followed by an oxidizing $H_2O$ pulse in each cycle. From an electrical standpoint, the optimization of the $ZrO_2$:$HfO_2$ cycle ratio was first carried out on a standard $SiO_2$/Si substrate, focusing on minimizing EOT and leakage current as a first-pass electrical screening. A 13:5 cycle ratio of $ZrO_2$ to $HfO_2$ was found to be optimal in our study. From materials standpoint, stabilization of NC behavior is closely tied to the formation of a mixed antiferroelectric–ferroelectric phase in HZO. The use of a bilayer structure allows for controlled modulation of crystallization pathways and phase stability during post-deposition processing. The in-situ deposition of a metal TiN capping layer by plasma-enhanced ALD also plays a crucial role by imparting confinement strain to the HZO, further promoting the stabilization of NC characteristics (*26*). We note that while the present study employed a 13:5 $ZrO_2$ :$HfO_2$ cycle ratio for integration into the N-polar GaN HEMT structure, the optimal cycle ratio and deposition process parameters may vary depending on the specific dielectric material system, underlying interface, and deposition environment (e.g., the Cambridge Fiji 200 Plasma ALD system used in this study).

We also note that the HZO/TiN stack in our work was deposited using standard ALD conditions and was not designed or optimized for strain engineering, unlike intentional high-stress capping layers such as $Si_3N_4$ (*49*). Additionally, as shown in **Fig. S10**, the extracted sheet resistance $R_s$ of the NC (with HZO/TiN) and Schottky (without any dielectric) devices is nearly identical, further indicating that the 2DEG density and mobility are unaffected. Therefore, we do not expect any significant mobility modulation due to mechanical stress from the gate stack.

**Electrical measurements.** The d.c. electrical measurements were performed with a commercial semiconductor parameter analyzer (Keysight B1505A Power Device Analyzer). 20-µm tungsten

(W) probe tips (Signatone SE-20TB) were used to establish contact within a commercial Signatone 1160 SERIES probe station. To minimize the influence of contact and lead resistances on the accuracy of the measured drain current, a four-probe measurement technique was utilized for the drain and source contacts. Capacitance ($C$)-voltage ($V$) measurements were performed using a commercial Semiconductor parameter analyzer (Keysight B1505A Power Device Analyzer) with a multi-frequency capacitance measuring unit (Keysight B1520, MF-CMU). Effective oxide thickness (EOT) simulations of Si MOS capacitors were performed using Synopsys TCAD simulation. The frequency-independent $C$-$V$ was fitted to the EOT simulations, in order to extract the EOT value of the gate stack.

**Materials characterization**. TEM, STEM, HRTEM and EDS analyses were performed at Covalent Metrology using a JEOL JEM-F200 Multi-Purpose Electron Microscope, operated at an accelerating voltage of 200 kV, and equipped with a Gatan OneView CCD camera. To prepare the samples, protective carbon, e-Pt, and i-Pt coatings were applied, and lamellae were extracted using the lift-out technique with a Thermo-Fisher (FEI) Helios UC FIB-SEM system. EDS spectra were acquired using a Dual JEOL JED-2300 Dry SDD EDS detector. AFM imaging was conducted on a Bruker Dimension Icon in tapping mode with an NSC-15 probe. SEM imaging was carried out using an FEI Nova NanoSEM 450 operating at 20 kV, equipped with an Everhart-Thornley detector (ETD) in secondary electron mode.

**HZO Phase analysis.** For the phase analysis of polycrystalline grains in the HZO layer, we used HRTEM images. The local interatomic spacing in the ultrathin HZO layer was measured from the raw HRTEM images using ImageJ software with the Bio-Formats and Tia-View plugins, through its line profile tool. Since only the heavy atoms (Hf/Zr) were clearly visible in the HRTEM images, for the phase analysis only the interatomic distances and the configuration of these heavy atoms were considered. The extracted lattice spacings were averaged across multiple periodicities. Several grains with distinct profiles and interatomic spacings (different from other known phases in HZO) were identified, allowing us to assign a phase to the corresponding grains. From the crystal structures, the d-spacings and the atomic configurations along different zone axes were extracted using SingleCrystal and CrystalMaker software in tandem.

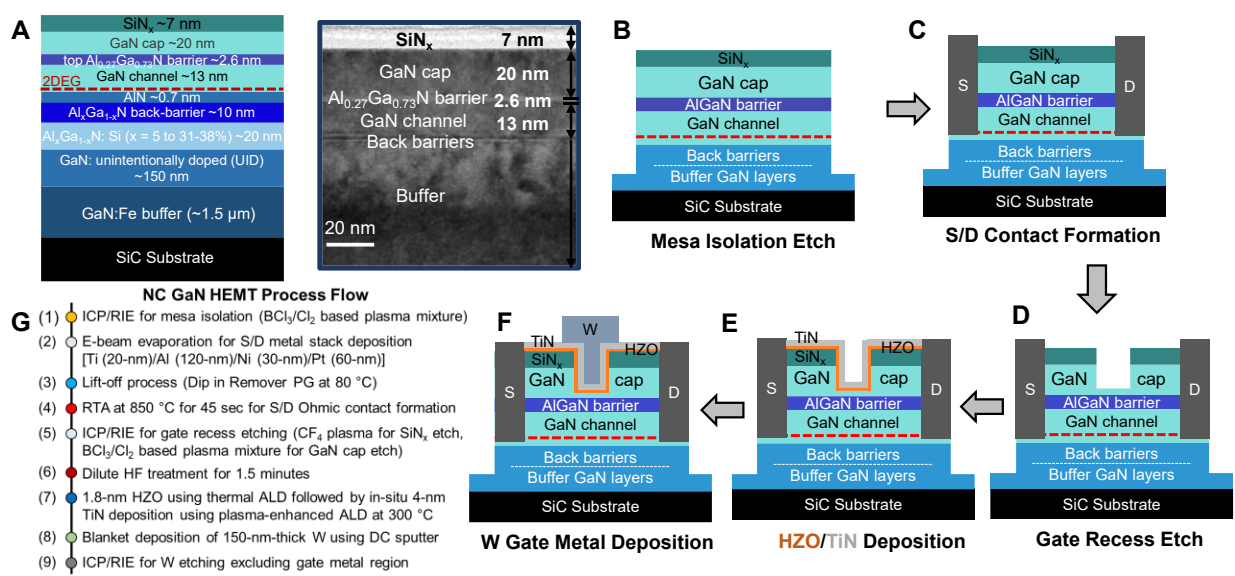

**Fig. S1. Schematic and device fabrication process flow. (A)** Schematic and TEM of the AlGaN/GaN heterostructure showing the individual layers and thicknesses in the stack. **(B-G)** Step-by-step fabrication process flows for the NC GaN HEMT. The fabrication process flow is further detailed in Supplementary Materials: **Materials and Methods** section.

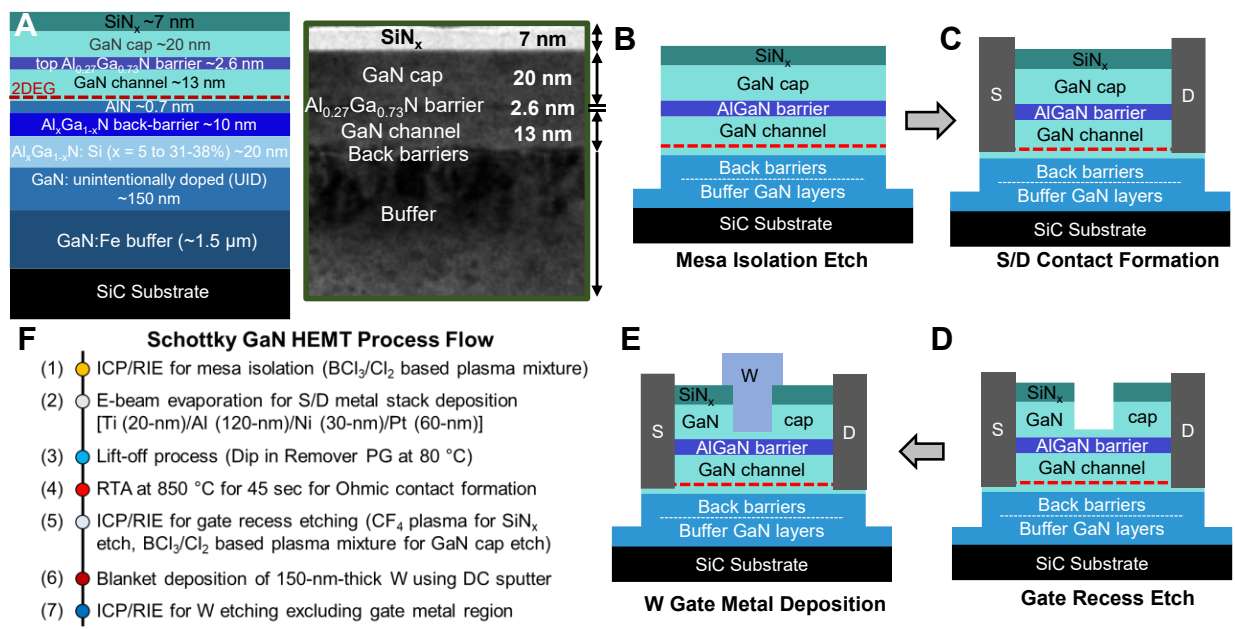

**Fig. S2. Schematic and device fabrication process flow. (A)** Schematic and corresponding TEM of the AlGaN/GaN heterostructure for Schottky HEMT showing the individual layers and their thicknesses in the stack. Note that this is identical to the heterostructure for NC HEMT (**fig. S1A**). **B-F**, Step-by-step fabrication process flows for the Schottky GaN HEMT.

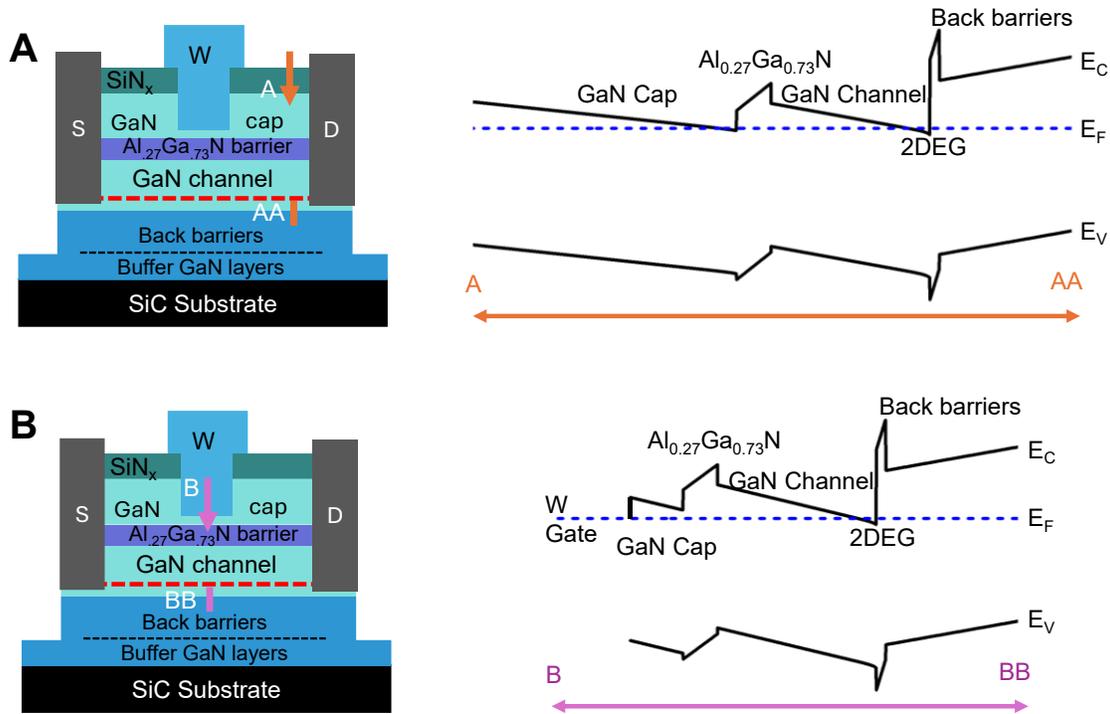

**Fig. S3. AlGaN/GaN heterostructure and formation of 2DEG for HEMT.** Energy band diagrams (**A,B**) of the conventional AlGaN/GaN heterostructures (top to bottom: GaN cap, $Al_{0.27}Ga_{0.73}N$ barrier, GaN channel, back barriers, thick buffer layers and SiC substrate) employed in this work leading to the formation of two-dimensional electron gas (2DEG) at the GaN channel and back barrier interface (shown in dashed red). The top $Al_{0.27}Ga_{0.73}N$ barrier layer induces a polarization-induced electric field that opposes gate current flow, while its wide bandgap reduces tunneling. The GaN channel provides a high-mobility transport medium for the 2DEG, ensuring efficient electron conduction. Below the channel, the wider-bandgap Al(Ga)N back-barrier layer induces the 2DEG channel and prevents electron leakage into the substrate for electron confinement. On top, the GaN capping layer acts as a passivation layer, reducing gate leakage, mitigating surface traps, and smoothing the electric field distribution to improve the 2DEG charge and device reliability. Together, these layers work synergistically to form and stabilize the 2DEG.

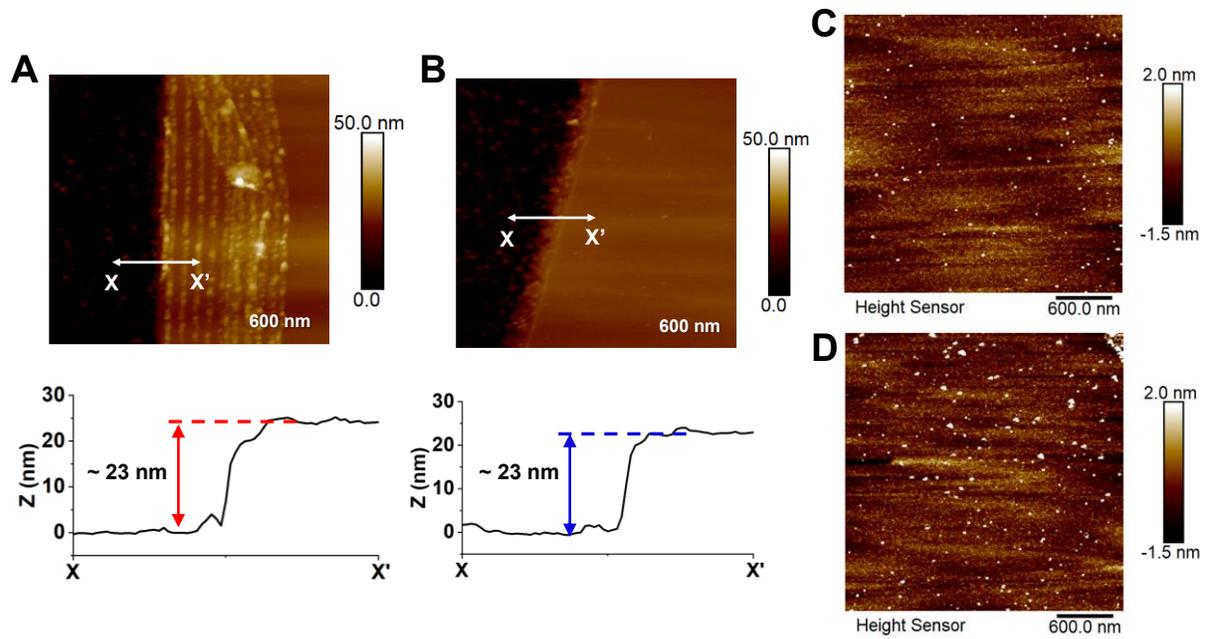

**Fig. S4. Atomic force microscope (AFM) images of gate recess etched region**. **(A)** AFM image of the gate recessed region for the NC GaN HEMT immediately after the recess etch, showing an etch height of ~23 nm (7 nm SiN$_x$ + 16 nm GaN etched), consistent with the target depth. **(B)** AFM image of the gate recess region for the Schottky GaN HEMT immediately after the recess etch, also showing an etch height of ~23 nm, confirming the identical gate recess profile for both devices. The identical remaining GaN cap thickness (~4 nm) after recess etch is further confirmed by the TEMs in **Fig. 2B,D**. We recall that both samples had 7 nm Si$_3$N$_x$ and 20 nm thick GaN cap (total ~27 nm) prior to any processing (**Fig. S1A, S2A**). Thus, the presence of ~4 nm remaining GaN cap layer in the TEM images (**Fig. 2B,D**) is further well-supported by the ~23 nm etch height in AFM measurements. Surface roughness of the recess-etched region of **(C)** an NC GaN HEMT, exhibiting a smooth surface with a root-mean square (RMS) roughness of ~0.62 nm over a 3 μm × 3 μm area; and **(D)** a Schottky GaN HEMT, exhibiting a similarly smooth surface with an RMS roughness of ~0.73 nm over a 3 μm × 3 μm area.

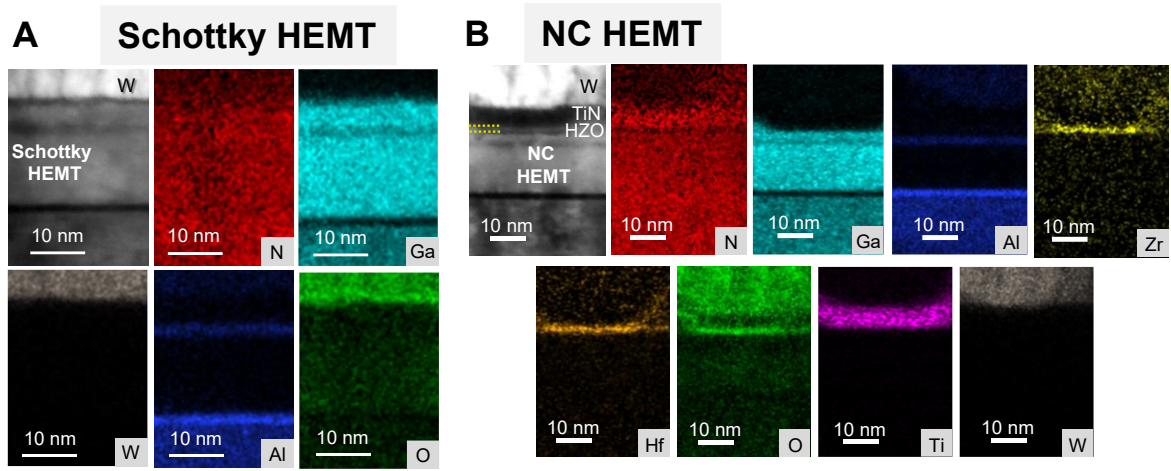

**Fig. S5. EDS characterization for Schottky and Ferroic-NC GaN HEMT structures. (A)** EDS mapping of the Schottky HEMT device layers, illustrating elemental distribution of N, Ga, Al, O, and W alongside corresponding scanning transmission electron microscope (STEM) cross-sectional image. **(B)** EDS mapping of the NC HEMT layers displays the elemental distribution of N, Ga, Al, Zr, Hf, O, Ti, and W along with the corresponding STEM cross-sectional image, further confirming the presence of the HZO dielectric layer.

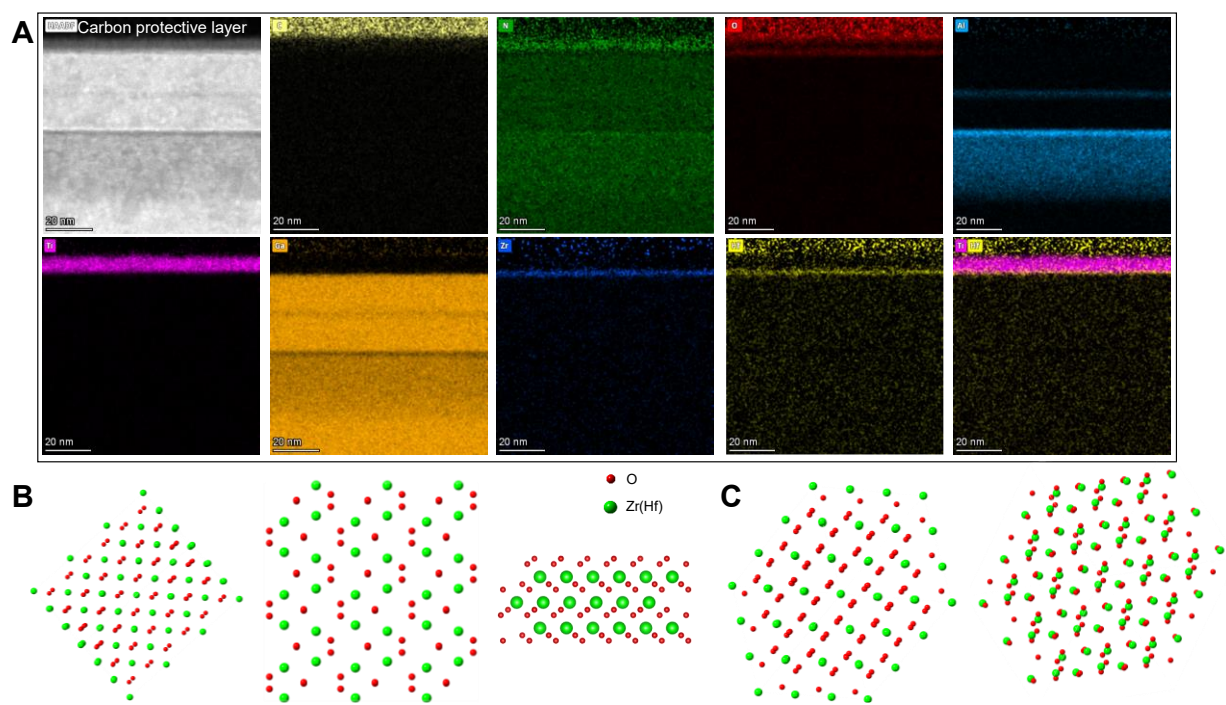

**Fig. S6. Additional TEM and EDS characterization of HZO on GaN and atomic configuration of HZO phases.** **(A)** Zoomed-out TEM and EDS analysis of the same stack as in Fig. **2E** and **2F**. EDS mapping displays the elemental distribution of N, Ga, Al, Zr, Hf, O, Ti, as well as the carbon protective layer used for imaging. **(B)** Atomic configurations of different phases with *d* spacing close to 2.64 Å (+/- 0.1 Å) (relevant to **Fig. 2E**). Tetragonal (P42/nmc, left) phase along [110] zone axis, monoclinic (p21/c, middle) phase along [002] zone axis, and orthorhombic (Pca21, right) phase along [100] zone axis. **(C)** Atomic configurations of different phases with *d* spacing close to 3.11 Å (+/- 0.1 Å) (relevant to **Fig. 2F**). Tetragonal (P42/nmc, left) phase along [111] zone axis, monoclinic (p21/c, right) phase along [11$\bar{1}$] zone axis. The orthorhombic (Pca21) phase doesn't have any d-spacing in 3.11 +/- 0.1 Å range. The crystal structures are obtained from Materials Project (*50*).

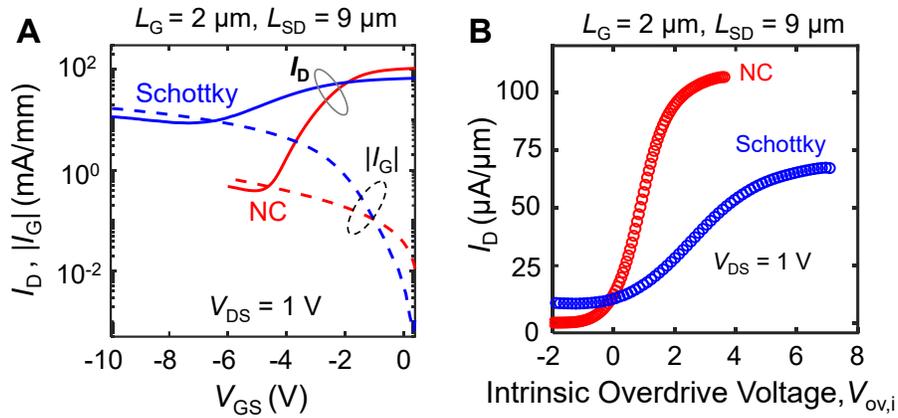

**Fig. S7. d.c. characteristics of additional HEMT devices. (A)** d.c. transfer characteristics ($I_D$ vs. $V_{GS}$) and gate leakage, and **(B)** $I_D$ vs. $V_{ov,i}$ (intrinsic overdrive voltage) for additional NC and Schottky HEMT devices both with $L_G = 2$ μm and $L_{SD} = 9$ μm. $I_D$ is normalized by channel width $W = 100$ μm and measured at drain voltage $V_{DS} = 1$ V.

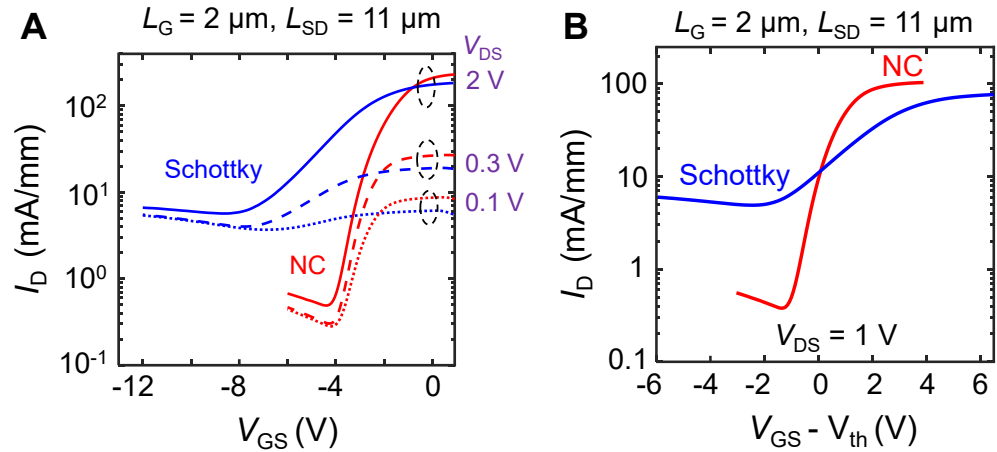

**Fig. S8. Additional d.c. transfer characteristics of HEMT devices. (A)** $I_D$ vs. $V_{GS}$ plots at various $V_{DS}$ (0.1 V, 0.3 V, and 2 V) display consistently higher drive current and lower leakage current for NC GaN HEMT compared to its Schottky counterpart. **(B)** $I_D$ vs. $V_{GS}$-$V_{th}$ (that is, $V_{GS}$ was adjusted by the threshold voltage $V_{th}$) at $V_{DS}$ = 1 V for the same HEMT in **A** (also in **Fig. 3C**). $L_G$ = 2 μm and $L_{SD}$ = 11 μm. $I_D$ is normalized by channel width $W$ = 100 μm.

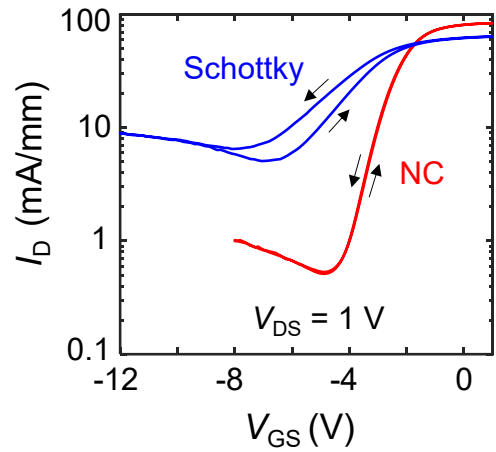

**Fig. S9. d.c. transfer characteristics and hysteresis in HEMT devices.** Bidirectional $I_D$–$V_{GS}$ characteristics for NC and Schottky GaN HEMTs. While hysteresis is observed in the Schottky HEMT, the NC HEMT exhibits nearly hysteresis-free behavior. Forward and backward sweeps are indicated by black arrows. $L_G = 2$ µm and $I_D$ is normalized by channel width $W = 100$ µm.

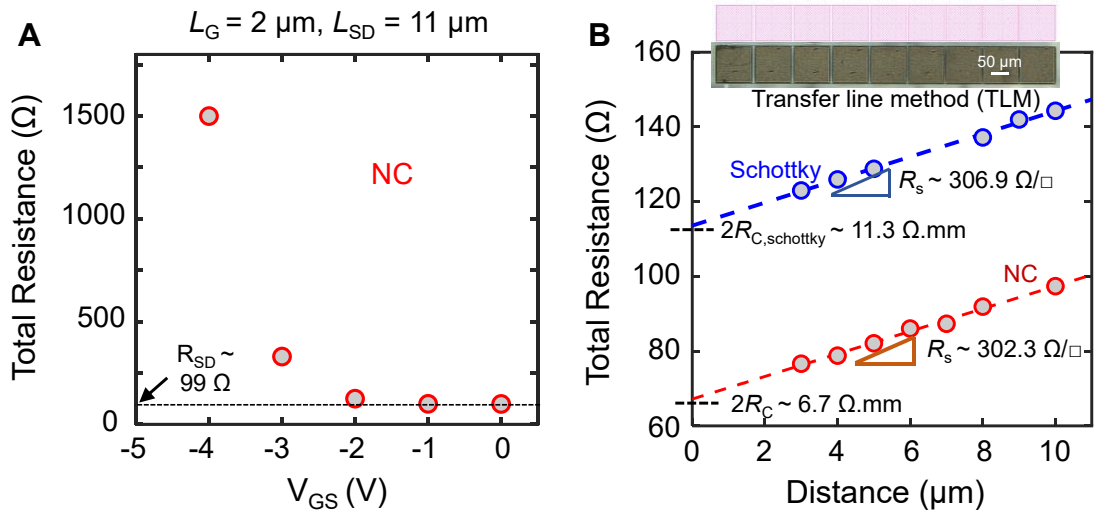

**Fig. S10. Series resistance extraction and TLM measurements. (A)** Measured total resistance of a representative Ferroic NC gated GaN HEMT ($L_G$ = 2 μm, $L_{SD}$ = 11 μm, channel width $W$ = 100 μm) as a function of $V_{GS}$ in the linear regime ($V_{DS} \leq$ 50 mV). At high $V_{GS}$, total resistance saturates, indicating that the channel resistance becomes negligible, and the resistance is dominated by $R_{SD}$. $R_{SD}$ is the sum of source resistance (source contact resistance + source access region resistance) and drain resistance (drain contact resistance + drain access region resistance). The extracted $R_{SD}$ for the NC gated HEMT is ~99 Ω (9.9 Ω.mm when normalized by $W$), as highlighted by the dashed line. $R_{SD}$ of both NC and Schottky HEMT devices are extracted using the approach described here. **(B)** Sheet resistance $R_s$ and contact resistance $R_c$ are extracted from the transmission line measurement (TLM) linear fit (dashed lines) of the total resistance vs. distance between the two consecutive contacts. The inset shows the mask layout and an optical image of the representative TLM structure used for the extraction with 4 probes used in the measurement. The contact spacings are in 1 μm increments. The extracted sheet resistance $R_s$ (from slope of the TLM linear fits), ~306.9 Ω/□ for the Schottky gate and ~302.3Ω/□ for the NC gated structures are comparable, indicating that the material quality and 2DEG charge density of the GaN channels in both devices are similar (the identical layer thicknesses of both stacks are confirmed by TEM in **Fig. 2** and **figs. S1A, S2A**).